**The interpretation of quantum mechanics and of probability:**

**Identical role of the 'observer'**

*Louis Vervoort*

*University of Montreal,*

louis.vervoort@umontreal.ca, louisvervoort@hotmail.com

17.06.2011

**Abstract**. The aim of the article is to argue that the interpretations of quantum mechanics and of probability are much closer than usually thought. Indeed, a detailed analysis of the concept of probability (within the standard frequency theory of R. von Mises) reveals that the latter concept always refers to an observing system. The enigmatic role of the observer in the Copenhagen interpretation therefore derives from a precise understanding of probability. Besides explaining several elements of the Copenhagen interpretation, our model also allows to reinterpret recent results from 'relational quantum mechanics', and to question the premises of the 'subjective approach to quantum probabilities'.

**1. Introduction.**

A key element of the Copenhagen interpretation of quantum mechanics is the role played by the observer, or rather the observing system. The observing system, or the measurement, makes the wave function collapse. By the same token it causes the 'measurement problem': why is an observing system any different from any 'normal' physical system, e.g. the natural environment - which leaves the wave function of the system in its superposition state ? Bohr and Heisenberg are reputed to be the first to have recognized the role of the observing system, giving an 'instrumentalist' or 'operationalist' flavor to quantum mechanics: to some, reality seemed to depend on, or be determined by, apparatuses. Since then a further shift in the interpretation of quantum mechanics has been proposed by several authors – a shift towards subjectivism, in which it is now the observer *as a human being*, including his or her mind, who plays the starry role: along with quantum information theory, Alice and Bob entered the scene. The degree of subjectivism is of course different for different authors, but the most radical of these interpretations (see Section 3.3.) almost impart the impression that quantum mechanics 'happens in the head of the subject', and leave the reader wondering where the objective basis of science is gone. Besides the standard Copenhagen interpretation, we will in the following investigate some of the better known new interpretations of quantum mechanics, namely 'relational quantum mechanics' of Rovelli and others [1-4], and the Bayesian or subjective interpretation of quantum probabilities of Bub, Caves, Fuchs, Schack and others [5-7]. As an excellent representative of the classical Copenhagen interpretation, we will use Peres [8-10], especially his textbook [8].

The aim of the present article is to show that the role of the 'observer' in quantum mechanics is not new: it is exactly the same as he/she/it plays in classical probability theory. More precisely, we will argue 1) that a precise definition of probability (à la von Mises) always refers to an observing system, 2) that (as a consequence) the instrumentalist aspects of the Copenhagen interpretation stem from the probabilistic nature of quantum phenomena, and 3) that also other interpretations of quantum mechanics [1-7] can be re-interpreted, less radically, within the standard interpretation of probability. The 'understanding' of quantum mechanics, beyond the formalism, would therefore heavily draw on the interpretation of the concept of probability.

Many a physicist will wonder whether anything new can be learned from the interpretation of probability. Is everything about probability not entirely said with Kolmogorov's simple axioms, dating from 1933 [11] ? Unfortunately not, else probability theory would not be termed the branch of mathematics in which it is easiest to make mistakes (Ref. [12], p. 4). Indeed, in order to apply probability calculus to the real world as it should one needs to know to *which type of events* exactly to apply it; in other words, one needs an interpretation of the concept of probability, beyond Kolmogorov's axioms. The most widespread interpretation in science is the relative frequency interpretation (in the limit of infinite trial series), which is generally attributed to Richard von Mises [13-14]. (References [13-14] offer, in our view, the most rigorous treatment of all aspects of probability theory, both foundations and calculus, and in particular their link.) However, other interpretations such as the classical interpretation of Laplace, the propensity interpretation of Popper, and the subjective interpretation, associating probability with 'degree of belief', exist (general references are [15-17] and the condensed [28] Ch. 4). As said, the subjective interpretation regains a vivid interest in the field of quantum mechanics [5-7].

As a matter of fact (and much to our own surprise), it appears that the notion of probability contains a few implicit notions, and that a minimalistic definition as P(R) = probability of result R = $\lim_{n \to \infty}$ {number of occurrences of R / total number (n) of trials}, does *not* guarantee flawless application, as we showed in Ref. [18]. The main result of [18] is simple: in order to calculate P(R) (with R a result, outcome, or event), that 'R' needs to be precisely specified, including the 'initializing' and 'probing' conditions in which it occurs. In other words, P(R) can be completely different from P(R') if R' is (at first glance) the same event as R but probed or measured in different circumstances (see Section 2). Let us stress that our conceptual study of probability is derived in a straightforward manner from von Mises' probability theory, which is explicitly objective. In view of the recent debate in quantum mechanics, we believe it is not superfluous to recall that not only von Mises but also other fathers of probability theory as Kolmogorov and his pupil Gnedenko, were well aware of, and actively engaged in, the objective – subjective debate (making it certainly more than 100 years old). They offered vigorous arguments in favor of an objective interpretation of probability (see e.g. Ref. [13] p. 75ff., 94ff., and Ref. [19] p. 26ff.). What did come as a surprise to us, is that a detailed conceptual study of the notion of probability appears so useful – it is an understatement to say that physicists are not used to such conceptual analyses. But we believe it is necessary to first explore in detail the physical and intuitive foundations of a concept as probability, before plunging into calculations. Else one may well derive results that look professional, but lead nowhere. To the best of our knowledge, the aspects we will highlight here have not been studied in detail elsewhere.

Before turning to the interpretation of quantum mechanics, we will in the next Section briefly recall the main results of Ref. [18]. Let us emphasize that our results may only be fully convincing if presented as a whole and in detail; we therefore ask the reader to consult [18] if necessary.

## 2. Results of the detailed interpretation of probability of Ref. [18].

Typical textbooks state that probability is attributed to 'random events'. When we want to experimentally determine the probability of the event e = 'this coin shows result R = heads', we all know what to do: we toss the coin many times and determine the relative frequency of heads (P(e) = P(R) = ½ for a regular coin). Now if we are looking for the *in-principle conditions on which P(e) depends* (these in-principle conditions will become essential in quantum mechanics), we quickly realize that these at least contain the initial and final conditions of the tosses: the way we start them and probe (observe) the result. Normal tossing involves a vigorous momentum to start with, and normal probing takes place, for instance, on a table. We could obtain *any* result for P(e) if we would put glue on the table and launch the coin with a sufficiently refined momentum distribution (with sufficient technological trickery we could moreover make the whole affair look perfectly random and regular if we wanted, but that isn't important here). *Therefore, in principle P(e) depends on the 'initializing' and 'probing' or 'observing' conditions in which e occurs.* (Equivalently, one could say that P(e) depends on the initializing and observing subsystem that form, together with the die, the probabilistic system, or 'p-system' [18]. In our example the initializing subsystem is typically a randomizing hand, and the observing subsystem a table.) In above example 'e' is a probabilistic event 'created' by a human; in such an artificial experiment we have no problem to conceive the role of probing or observing conditions. Now does the above mentioned dependence also hold for *natural* probabilistic phenomena (diffusion, collisions, quantum phenomena,…) that happen every instant everywhere in the universe without anybody observing ? In [18] we argued that the role of the initiating and probing conditions is taken here by the *environment* (temperature, force fields etc. determine the probabilities in question in an obvious way). At any rate, when we *verify* or *measure* the probabilities which e.g. quantum theory predicts for an event occurring in a specified environment, we need to do the experiment in that same environment or in a lab reproducing the environment – thus the initiating and probing conditions reappear. Therefore, it is safe to generalize above simple idea [18]: for calculating or determining a probability, as soon as things get somewhat subtle, one should remember the in-principle dependence on the initiating and probing conditions (or the environment if one prefers). (Ref. [18] proposes a detailed definition of probability that includes these conditions[1].)

However simple the idea, inspection of the countless paradoxes which the fascinating topic of probability has to offer, left us little doubt that it is this idea which is at the basis of *all* these paradoxes. Von Mises exposes many tens of them in his works [13-14]; many textbooks contain at least a few. Let us look at a few well-known examples. Bertrand's paradox (dating from the end of the 19th century) goes as follows: "A chord is drawn randomly in a circle. What is the probability that it is shorter than the side of the inscribed equilateral triangle ?" Since a chord can be drawn randomly in a variety of ways [22] (as may not be obvious at first reading), the answer is simply undefined. Each method of random drawing (read: each initializing condition of the random experiment) leads to a specific probability. In other words: to have a defined probability, one needs a defined initializing condition. The infamous 'Monty Hall' problem [23], amply discussed and popularized in the nineties in all media by Marilyn vos Savant, will be seen to become almost trivial if one remembers that, here to, one looks for a relative frequency of an outcome R of a well-defined experiment[2].

Essential for the following is the role of the probing or observing conditions. Any numerical value of *any* probability depends, in principle, on the observing conditions, as was already suggested by the coin example [18]. In a macroscopic and thoroughly known case as a coin toss, when asking a person to determine 'the probability of heads for this coin', no-one feels compelled to specify these conditions. Everyone assumes that everyone knows what is meant with the event 'e'. But in the quantum world we don't know much anymore; as we will see further the conditions become essential.

However, on further inspection they play a crucial role too in classical systems, namely to safeguard the objective nature of probability. A question that almost automatically pops up in discussions on probability, is: "but doesn't probability depend on our knowledge ?" (It is this question that opens the door to subjective interpretations of probability.) Consider for instance a regular die throw. 'For me', the probability that a 6 shows up is $1/6^{th}$, but 'for Alice', who is equipped with a sophisticated camera and can register the die movement in real-time until it comes to a halt, that same probability seems to be either 0 or 1. Or, 'for me' the chance that I will catch a bus during the next 10 minutes is 10/60 (I know that one bus passes per hour). But someone who sits in a tower and has a lookout over the city traffic might disagree [18].

---

[1] Note that in this definition the initiating conditions, the observing conditions, and the environment intervene in an exactly symmetric manner.
[2] R = 'I win a car by switching'; P(R) = 2/3.

It is not difficult to devise paradoxical situations as these, but they stem from a neglect of the fact that P(R) is only well-defined – only exists – if R is well-defined, including its probing conditions. Doing an experiment in which a die is thrown and probed in the usual way is one thing; throwing and then monitoring it in real-time with a camera is another. These two events are different due to different observing conditions (or observing subsystems); for *that* reason the probabilities are different, not because of two different knowledge states or degrees of belief (see Appendix for a detailed argument).

Now, as we show in some detail in the Appendix, the subjective shift is extremely tempting. Indeed, it is obvious that in above die example one could introduce *conditional* probabilities (e.g. the probability that the upper surface shows 6 at halt *if it is given that* the surface shows 6 one μsec before the die comes to a halt as observed with a camera). *Any* probability can in principle be considered a conditional probability (see Kolmogorov, Ref. [11], p. 3; Ref. [19] p. 67). And the expression "if it is given that" seems to be equivalent, in this context, to "if it is known that", thus invoking the knowledge state of an observer in a subliminal manner. This may be the origin of why subjective interpretations are so popular, as we argue in the Appendix.

If we assume that probability theory is part of science, then it should necessarily concern 'objects' that can be separated from the 'subject' (we): if our minds would play any role, i.e. determine the objects (= probabilities) under study, scientific, i.e. 'objective' comparison of results would be impossible (in the context of quantum mechanics, this idea has been analyzed and stressed by Bohr at several occasions). In [18] we showed that if one wants to see probability as an objective measure, there is a simple way to do so: attribute it to events under defined probing conditions. True, 'objective' (and 'observer-independent' even more) is a somewhat tricky word here: it means '*identical (and mind-independent) for all observers performing identical experiments*', so *objective in the scientific sense* – even if the observer, or rather the observing subsystem, is in a sense part of the system ! (Remember that the observing subsystem is part of the p-system.) It is therefore clear that in objective accounts 'observer' must be understood as a physical system without a mind. The subjective shift is equivalent to including a human mind in the 'observer'.

A last remark that will prove useful is the following. In [18] we argued, following authors as von Mises, Popper [20], and van Fraassen ([21], Ch. 6) that it is often helpful to remember that probability belongs to 'experiments', rather than just events: probability theory concerns repeatable and real physical tests or experiments. The 'event space' of axiomatic probability theory [11] is a vague concept[3]. Probability theory is not only a mathematical theory: it applies to real-world events occurring under repeatable conditions.

Let us now see how these findings compare to the interpretation of quantum mechanics.

## 3. Conceptual link between probability and quantum mechanics.

### 3.1. Objective probability and the Copenhagen interpretation.

In view of what we learned above, it seems that the importance of the 'observer' in quantum mechanics should not surprise us, if quantum phenomena are probabilistic phenomena. One of the key ideas of the Copenhagen interpretation is expressed in following quote by Bohr, taken from his 1935 reply to Einstein, Podolsky and Rosen in their debate on the completeness of quantum mechanics. "The procedure of measurement has an essential influence on the conditions on which the very definition of the physical quantities in question rests" [25]. Is this passage understandable ? (we did not understand it before our investigation [18]). In Ref. [24] Bell not only identifies above passage as the essential ingredient of Bohr's reply, but also admits and details his complete incapacity to understand it (p. 58). However, within the objective interpretation of probability of Section 2, it seems we can simply re-interpret what Bohr claims. It appears now that Bohr says nothing else than that quantum properties *are determined by the observing conditions or in other words the observing subsystem[4]*. But we found in former Section that this dependence holds – in principle - for all probabilistic systems, quantum or classical. For classical systems one needs some attention to find examples of the influence of the observing system; in quantum mechanics the 'in principle' becomes basic - as correctly emphasized by Bohr. As an example: the probability that an x-polarized photon passes a y-polarizer obviously depends on x and y. *Any* numerical probability value, quantum or classical, is a function of the analyzer / detector parameters.

---

[3] It is of course mathematically well-defined within a system as Kolmogorov's; but it does not say to which events probability theory can be applied. Many or most 'events' are not probabilistic.
[4] This 'determination' can of course only materialize through an interaction between the trial system and the observer system (think of how any classical probability is measured). This idea is elaborated by Bohr in Ref. [25] by invoking the 'quantum of interaction'.

Seen from this angle, one would think that the 'measurement problem' becomes, in essence, transparent. The collapse of the wave function (the selection of one of the outcomes of the experiment that is encoded in the wave function), strongly reminds us, of course, of classical probabilistic measurements in which the observing subsystem selects one possibility (e. g. 'six up' in a die throw selected by a table). We believe the concept of collapse contains in essence no more mystery than the 'determination' of a classical probabilistic system during the act of observation. (This does not preclude that more refined mechanisms might underlie the simple probabilistic manifestation that emerges from these mechanisms.)

Let us have a look at a few salient statements of Peres in Ref. [8]. "*A state is characterized by the probabilities of the various outcomes of every conceivable test*" ([8] p. 24, where these 'tests' should be understood as pertaining to an experimental set-up). A little further: "Note that the word 'state' does not refer to the photon by itself, but to an entire experimental setup involving macroscopic instruments. This point was emphasized by Bohr […]" ([8], p. 25). We are by now inclined to say that these quotes hold for *any* probabilistic system, quantum or classical. To a die throw one could obviously associate a 'state' consisting of the six discrete results and their probabilities. At any rate, also the probability of an outcome of a die throw depends crucially on the 'entire experimental setup' that is used to perform it, in particular the initial and probing conditions (Section 2). Probabilities, and in particular quantum probabilities, are measures pertaining to repeatable experiments, *not* objects per se [18]. Compare to following passage of [8] (p. 73): "The notion of density matrix – just as that of state vector – describes a *preparation procedure*; or, if you prefer, it describes an ensemble of quantum systems, whose statistical properties correspond to the given preparation procedure". In Section 2 we termed the 'preparation' of this quote the 'initializing' of the probabilistic system. We can review here only a few of the instrumentalist passages of [8], but it appears they all can be interpreted by using the same notions as probability theory already uses, explicitly or implicitly.

Note that we do not claim that every element of the Copenhagen interpretation has its counterpart in probability theory (à la von Mises). Notably, the latter has no uncertainty relations, neither the corresponding commutation relations. But interestingly enough, even these seem to be foreshadowed by an adequate interpretation of probability. The commutation relations stipulate, among other things, which observables A and B can be measured simultaneously. Now, within von Mises' framework, P(A&B), the joint probability of A and B, is defined *only if A and B can be measured by an experimental set-up that allows to measure A, B, and A and B (combined) by using the same observing system*[5]. In symbols one has: $P_C(A.B) = P_C(A).P_C(B|A)$, with C the *same* experimental conditions in the *three* experiments. One could equally well write $P(A.B|C) = P(A|C).P(B|A.C)$: see the Appendix. In other words, within von Mises' probability theory the question of simultaneous measurement of two random quantities is already crucial – just as it is crucial in quantum mechanics. Von Mises warns explicitly that not of all quantities A and B the joint distribution exists; it is not difficult to find conditions C in which A and B cannot be simultaneously measured. We see here again that (via the commutation relations) quantum mechanics fills in, in a stringent manner, the conditions under which to apply classical probability theory in the atomic and subatomic realm[6].

Let us note, finally, that probability theory is less authoritative than the Copenhagen orthodoxy. On the Copenhagen interpretation, the commutation relations stipulate not only which observables can be measured simultaneously, but also which observables *exist* simultaneously. Note that this is indeed *the second essential ingredient of Bohr's answer to EPR*, besides above passage. (As we read it, Bohr's argument can thus be summarized as follows: measurement brings observables into being through an inevitable interaction with an observing system; if two observables cannot be measured simultaneously, they do not exist simultaneously.) Probability theory is of course silent about the existence of outcomes when they are not measured (i.e. about their being determined by hidden variables, whether measured or not). Laplace and Einstein famously argued that probabilities hide causal mechanisms - an option that is fully left open by probability theory, but prohibited by the Copenhagen interpretation.

---

[5] This follows from von Mises' exposition in Ref. [14] pp. 26 – 39, even if it is not entirely explicit. The key procedure is the following: start from two collectives (experimental series), one in which A is measured, one for B. If A and B are 'combinable', then one can construct a collective for the joint measurement of A and B, and determine their joint probability. On a straightforward interpretation of von Mises, the latter collective must correspond to an experimental series *using the same equipment as used for measuring A and B separately*. (Thus a 'joint' measurement is a 'simultaneous' measurement.) Von Mises warns explicitly that not just of any A and B the joint probability can be measured.

[6] At this point it is tempting to suggest a link with the notorious quantum non-locality or contextuality revealed, e.g., by the theorems of Bell and Kochen-Specker. Indeed, we argued that a typical quantum correlation as P(A.B) has to be understood as P(A.B|C) and thus depends *in principle* on C, i.e. on all the parameters of the set-up, for instance polarizer or magnet directions. Quantum mechanics shows that the 'in principle' should be taken seriously: quantum probabilities are 'contextual', i.e. do depend on the whole experimental context – as already suggested by the notion of probability. But we cannot elaborate this point here, and leave the difficult question whether this is 'all there is to quantum non-locality' open.

**3.2. Objective Probability and Relational Quantum Mechanics.**

An interesting interpretation of quantum mechanics is due to Rovelli [1-4], who termed it 'relational quantum mechanics' (RQM). The essential idea of this account is to consider any state vector as relative to an observer, or rather observing system (for Rovelli anything can be an 'observer', e.g. an atom). A relevant quote is this: "The notion rejected here is the notion of absolute, or observer-independent, state of a system; equivalently, the notion of observer-independent values of physical quantities. The thesis of the present work is that by abandoning such a notion (in favor of the weaker notion of state – and values of physical quantities - *relative* to something), quantum mechanics makes much more sense" [1] (Quote 1). Just as Einstein's rejection of the obsolete notions of absolute time and simultaneity allowed to reinterpret the Lorentz transformations, Rovelli conjectures that the replacement of absolute states by relative states allows to found quantum mechanics on more solid grounds [1].

Now, according to classic quantum theory [8], a quantum state represents outcomes of an experiment and their probabilities (see Section 3.1.). But we have argued in Section 2 that these probabilities necessarily depend on, e.g., the observing subsystem; therefore, also the corresponding outcomes necessarily depend on the observing subsystem[7]. But this seems to offer an explanation of the main claim of relational quantum mechanics (Quote 1), stating that quantum states and the physical quantities they represent are *relative* to an observer. They indeed are: they depend on it, or are determined by it. Therefore, our objective probability interpretation of quantum mechanics provides a natural basis to understand the relational nature of quantum mechanics.

---

[7] For instance, a certain judiciously constructed die throw experiment, containing as probing system a table covered with glue, may have as unique outcome R = 6, with probability 1. With another probing system (remove the glue) the outcomes are 1,…,6 with probability 1/6. Probabilities *and* outcomes depend, in principle, on the probing.

In Ref. [2] the authors analyze the EPR paradox. The 'solution' (or rather interpretation) of this paradox within RQM "consists in acknowledging that different observers can give different accounts of the actuality of the same physical property" [2]. Let us have a closer look at the argument (as we read it). According to RQM, the state vector ψ of a system S is relative to observer A (Alice), in the sense that ψ represents a coding of the outcome of previous interactions between S and A. Now, crucially, these interactions "are *actual only with respect to A*" (Quote 2, our italics); "the state ψ is only relative to A: ψ is the coding of the information that A has about S" [2]. In the EPR gedankenexperiment, involving a spin (σ) singlet state for a pair of electrons I and II, Alice may measure at $t_0$ e.g. $\sigma_{z,I} = +1$; then she knows that Bob necessarily will find at times $t \geq t_0$ (time relative to her), that $\sigma_{z,II} = -1$ (similar for $\sigma_{x,I}$ and $\sigma_{x,II}$)[8]. Nevertheless, at $t_0$ $\sigma_{z,II}$ *has no reality for Alice* (there is no element of reality corresponding to $\sigma_{z,II}$ for Alice): see Quote 2, or "recall that a property of S is actual relative to A only if substantiated in a correlation between A and S" [2]. Therefore, the EPR argument brakes down from the start (it's not even necessary to consider the second part, involving $\sigma_{x,II}$). Since there is nothing real happening with II for Alice (only her subjective knowledge state about II changes), a fortiori nothing real is transmitted, so as a bonus RQM saves locality [2].

The key argument is therefore that, in the case of quantum properties, it makes no sense that A talks about a property that is not instantiated by an interaction with her: *it does not exist*. This is a radical stipulation; even if some quantum results (Bell's theorem) hint into this direction, it remains a matter of taste whether one wants to accept it or not. Ref. [2] is of course not intended to be a logical refutation of EPR within a generally accepted theory. EPR may equally well remain on their position and emphasize that the perfect correlation between $\sigma_{z,I}$ and $\sigma_{z,II}$, and the fact that Alice can predict with certainty what Bob will measure, remain serious arguments for the simultaneous existence of both quantities (for all observers). True, Bell's theorem has made, for many physicists, such a claim much less plausible; but we believe much remains to be said about the famous theorem [8, 24, 26, 30].

---

[8] In the notation of RQM properties should, for obvious reasons, always be indexed to refer to the observer, so the authors of [2] rather use symbols as $\sigma_{z,I}^A$. But we can simplify here without loss of information.

There is one radical aspect of RQM that can be re-interpreted within our classical probabilistic interpretation in a less drastic manner. Indeed, it is not fully clear to us to which observer the singlet state

$$\Psi = (1/\sqrt{2}) \{ |u_{I,+}\rangle \cdot |u_{II,-}\rangle - |u_{I,-}\rangle \cdot |u_{II,+}\rangle \}$$

refers in RQM, since it contains information relative to two observers according to that interpretation. In [2] the authors show that RQM is compatible with the classic quantum result encoded in $\Psi$, namely that for all electron pairs $\sigma_{z,I} = -\sigma_{z,II}$ (their argument is refined in [4]), but they reject the idea that $\Psi$ could be relative to a 'superobserver' [2] or to 'the Lorentz frame that we arbitrarily choose to perform our calculations' [10] (Refs. [10] and [29] accept such an observer).

If we interpret quantum mechanics within an objective probability approach, we believe one should not have such scruples vis-à-vis a superobserver. Indeed, as stated above, within von Mises' framework the joint probability P(A&B) is defined if A and B can be measured by an experimental set-up that allows to measure A, B, and A and B (combined) by using the same observing system. Therefore, classical probability theory suggests that the superobserver in the EPR experiment is simply Cecil, the observer who can measure $\sigma_{z,I}$ and $\sigma_{z,II}$ simultaneously. *Such an observer manifestly exists* (cf. the well-known coincidence measurements done to verify Bell's theorem[9]); the natural interpretation seems to be that $\Psi$ is relative to Cecil. But $\Psi$ can also be used as a theoretical tool by Alice and Bob. Classical probability theory suggests that what really matters is the experimental set-up used (*this* defines the 'observer'), not really where the human being using the apparatus or calculating the probabilities in question 'sits'. Indeed, probability theory allows to predict and use single-event probabilities such as P(A) and P(B), corresponding to the measurement of just A or B (the observer measures one quantity by using one part of the observing system), but also of double-event or joint probabilities such as P(A&B) (the observer measures both): the 'magic' is that single and double-event probabilities are mutually linked by objective probabilistic laws (or regularities). On our view, this is in perfect harmony with quantum mechanics (which seems a relief). The information coded in $\Psi$ reveals such an objective probabilistic regularity that is expressed as a function of single-event states. Analogously as in the classical probabilistic case, the joint probabilities coded in $\Psi$ (such as P(+&-) = P(-&+) = 1/2) exist *for*

---

[9] The existence of such a simultaneous measurement is allowed by a vanishing commutator $[\sigma_{z,I}, \sigma_{z,II}]$.

*the same apparatus* as the single-event probabilities (coded in the $|u_{i,j}\rangle$), even if, so to speak, Cecil uses both arms of the EPR experimental set-up, and Alice and Bob use only one.

Therefore, however interesting we find RQM, we could relax such statements as "[…] we can never juxtapose properties relative to different systems" [2]; or in any case we should treat this phrasing with much care. What is meant is that $\sigma_{z,I}$ and $\sigma_{z,II}$ cannot be juxtaposed since relative to different observers. We believe the better reading is that $\sigma_{z,I}$ and $\sigma_{z,II}$ *can* be juxtaposed[10] *because relative to one and the same apparatus*, so relative to one real observer (Cecil, the apparatus). In sum, science sometimes allows to predict probabilistic regularities for double events. These double-event probabilities necessarily involve a joint observing system (e.g. a table on which two dice can be observed, or two arms in an EPR experiment). If these objective regularities predict $P(+\&+) = P(-\&-) = 0$, as in the EPR case, then we (Alice, Bob, Cecil - the humans) know that if $\sigma_{z,I} = +1$, $\sigma_{z,II}$ necessarily $= -1$. Note that our interpretation in reality only deals with 'observing systems'.

### 3.3. Von Mises Probability and Subjective Quantum Probability.

The subjective interpretation of quantum probability is vigorously defended in recent Refs. as [5-7]. Let us remark from the beginning that we do not intend to scrutinize the many surprising statements these articles contain. We indeed believe, in view of the centennial history of the subjective – objective debate in probability theory, that it may well be impossible to convert adherents of the subjectivist interpretation to an objectivist position (and v.v.): the premises seem too different; they may ultimately be metaphysical; and identical facts can be interpreted in different ways. We will therefore not venture into a detailed analysis of [5-7], but just relate a few statements of [7] to our arguments of the preceding Sections.

Let us first recall that strong arguments against the subjective interpretation were already offered by the fathers of probability theory (Section 1). More recently, the latter interpretation has been incisively criticized in the case of applications in medicine, criminology etc. [27], and in general [28, Ch. 4]. One of the main ideas of von Mises (Ref. [13] pp. 96-97) and Refs. [27-28] is that it is extremely dangerous to attribute probability to propositions (except if the link with experiments is unambiguous): the danger of running into unscientific interpretations is huge[11]. The main question that remains unanswered within the subjective approach is, it seems: if probabilities are just subjective degrees of belief of some observer, why do observers all over the world measure the same (quantum) probabilities when they do the same experiments ? Why do these different observers measure the same electron energies in solids, the same wavelengths emitted from atoms, the same EPR state probabilities, etc. ? Note that *all* probabilistic predictions of quantum mechanics are measured or verified by determining relative frequencies, as any experimental physicist will testify.

---

[10] Operationally by Cecil if one likes, but even theoretically by Alice and Bob.

The only point of [7] we will briefly comment here is the following. In order to justify and substantiate any difference between the subjective interpretation and the usual frequency interpretation, the authors (but this remark holds mutatis mutandis for all subjective approaches) have to invoke such a highly questionable premise as Lewis' 'principal principle' [7], distinguishing objective chance and (Bayesian) probability. Bayesian (real) probability (Pr) should satisfy, according to this principle: Pr(E C&D) = q, with E an event, C the proposition "the objective chance of E is q" (q *is* supposed to be the objective chance (!)), and D "some other compatible event, e.g., frequency data" [7]. Even if we make abstraction of the fact that the probability function 'Pr' has as argument an improbable mixture of events, propositions and data, Lewis' principle sounds remarkably circular. What is the use of introducing Pr if it is determined by q ? Why not stick to the objective measure q, which the authors seem to interpret themselves as von Mises' relative frequency ?

Now, we do not deny that there is a grain of truth in the subjective interpretation. For instance when we read: "This approach [the subjective interpretation] underlines the central role of the agent, or observer, in the very formulation of quantum mechanics" [7]. We have emphasized the role of the observer – *as physical system* – throughout this text. But we have also argued that by attributing probability to experiments, or to composed events including observing conditions, probability can be defined in a fully objective manner. True, the subjective shift is tempting: in order to make the concept objective, one has to include the observing *subsystem* (*not* the mental state of a human observer) into the probabilistic system. To make things worse, any probability is in reality a conditional probability; and conditional probabilities are almost always read 'if it is known that' (see Appendix) – suggesting a link with belief states. But probability theory and quantum mechanics are not psychological theories, but theories about repeatable *and therefore* objective experiments: if we would not find the same relative frequencies in the same repeated conditions, we would stop doing quantum mechanics and statistics.

---

[11] For instance, talking about the 'probability' of the propositions p = 'Theory X is correct' or p = 'Person X is guilty' makes no sense. Which experiments would have to be performed in order to measure P(p) ?

**4. Conclusion.**

The aim of the present article was to show that the interpretations of probability and of quantum mechanics are more overlapping than usually thought. The bridge between the two is the essential and identical role that the 'observer' plays in both frameworks. Indeed, a detailed conceptual study of probability (within von Mises' theory) reveals that any numerical value of a probability (P(R)) depends, is determined by, the initial and probing conditions in which R occurs. Therefore, any scientific probability is only defined if the observing subsystem is defined; it only exists 'relative' to such an observer system. In a natural environment, when nobody looks, the role of the 'observer' is played by the precise physical conditions imposed by the environment (the latter determine the probabilities in exactly the same manner as the conditions imposed by a human observer in a laboratory test; more precisely, they intervene in an exactly symmetric manner in a precise definition of probability [18]). This simple observation allows to interpret many elements of the Copenhagen interpretation: the 'understandable' (and measurable !) ingredient of the Hilbert space formalism are probabilities – and these should satisfy the above mentioned dependence on the observing system. In classical probabilistic systems this dependence is rarely or never explicitly mentioned (everyone knows how to perform and probe a regular die throw, no need to mention the exact conditions). But quantum theory has shown that in the quantum realm these conditions are stringent and essential.

Besides revisiting some classical elements of the Copenhagen interpretation [8], we could derive the essential claim of relational quantum mechanics [1-4]; relax one of its more radical conclusions [2]; and criticize the main premise of the subjective approach to quantum mechanics [5-7]. These findings seem to clearly indicate, we believe, that quantum mechanics, just as probability theory, is exclusively concerned with interactions between inanimate test systems and observer systems; these systems mimicking nature to some degree, also when we don't look.

Acknowledgements. We would like to acknowledge valuable comments of participants at the Conference of the Canadian Society for the History and Philosophy of Science in Montreal (2010). For detailed discussion of the issues presented here we would like to thank Yvon Gauthier, Michael Hall, and Jean-Pierre Marquis. Discussions on concrete examples with Mario Bunge and Vesselin Petkov motivated us to start the present work, and sparked essential ideas (which does not necessarily mean they will agree with all aspects of the framework).

**Appendix. Objective probability and the subjective shift.**

Let us have a more detailed look at the example of the die throw in Section 2. One may consider the probability of event $e_1$ = "the outcome R = 6 in a regular die throw (probed as usual on a table)". One could also consider the probability of $e_2$ = "R = 6 after a camera has registered that R = 6 one microsecond before the die came to a halt". The probability of $e_1$ = $P(e_1)$ = 1/6, while $P(e_2)$ = 1 (or suppose so). These probabilities are *not* different because of different subjective knowledge states. First note that also for $e_2$ one can define the probability without referring to 'subjective knowledge' or 'information'; $P(e_2)$ could be measured by an automat[12]. So $P(e_1) \neq P(e_2)$, not because someone has a different strength of belief, but because $e_1 \neq e_2$: both probabilities concern different events, different experiments, and in particular different observing conditions. This is a simple consequence of the basic idea of von Mises' theory: every probability can be seen as the result of a (series of) experimental tests.

Now since nothing is simple in the foundations of probability, it is at the same time true, as is obvious for any practitioner of probability calculus, that $P(e_2)$ can also be expressed in a manner that invokes, implicitly or explicitly, the status of knowledge of some observer – whence a potential confusion. Indeed, $P(e_2)$ can be considered a conditional probability, namely the probability that R = 6 *if it is given* that R = 6 one μsec before halting; in symbols $P(e_2) = P(e_1 \mid R = 6$ one μsec before halting). *Now it indeed seems that the phrasing "if it is given that" is equivalent in this context to "if it is known that"*; and therefore one is tempted to interpret conditional probability as depending on the knowledge of some observer. According to our point of view, this is fine as long as one realizes that this interpretation is a shortcut of thought, opening the door to subjective interpretations – it is in any case by no means the only interpretation. As we just illustrated, and as is explicitly proven by von Mises[13], conditional probabilities can very well be seen as corresponding to series of automated experiments in which no intervention nor belief state of a human agent is necessary. We cannot stress enough the fact that one can immediately derive from von Mises' treatment that such a non-anthropocentric interpretation is valid for *all* conditional probabilities: they all can be seen as describing objective experiments (whether related to chance games as in the above case, or to natural phenomena). But is this not a happy argument for the homogeneity of probabilistic phenomena ? Under the usual interpretation all *natural* stochastic phenomena occur according to probabilistic laws also without a human pondering about them; therefore the same should hold for *artificial* chance phenomena such as die throws[14]. It is worth to emphasize that conditional probabilities can be interpreted in an objective way, since they play an essential role in probability theory. Indeed, as already implicit in Kolmogorov ([11], p. 3), and explicitly stated by Gnedenko ([19], p. 67), *even so-called unconditional probabilities can be regarded as conditional on the circumstances (conditions) of realization*. Note that we stressed throughout Ref. [18] that it is useful to remember that these conditions are in principle composed. Therefore the natural generalization in our model is to replace, if helpful, P(R) by P(R|C) where C contains all relevant parameters that describe the initial, final, and 'environmental' experimental conditions.

---

[12] Such an automated experiment is this: let a robot launch a die, let it select by camera vision those trials that show R = 6 one μsec before the die comes to a halt, and let it measure on that ensemble R again at full stop of the die (all this could be done by a machine). The relative frequency of these results will converge to 1, as our robot could determine.

---

[13] More precisely, one can immediately derive from von Mises (Ref. [14], pp. 22-24) that the usual generic expression P(A|B) is the probability that event (result) A occurs *in experiments in which also event (result) B occurs*. All probabilistic concepts, such as conditional probability, are defined by von Mises by using collectives – i.e., in short, series of experiments.

[14] Van Fraassen (Ref. [21], p. 164ff.) gives a more detailed analysis of how subjective interpretations (probability linked to ignorance) originating from chance games slip into the objective statistics of physics. Also Gnedenko gives an enlightening analysis of the subjective shift ([19], p. 26ff.).